\documentstyle[pra,aps,floats,epsfig,12pt]{revtex}
\tightenlines

\def\bS{{\bf S}}
\def\bR{{\bf R}}
\def\bQ{{\bf Q}}

\begin{document}
\title{The no-sticking effect in ultra-cold collisions}

\author{Areez Mody, Michael Haggerty and Eric J. Heller}

\date{August 2000}

\address{Department of Physics, Harvard University, Cambridge, MA 02138}

\maketitle

\begin{abstract}
We provide the theoretical basis for understanding the phenomenon in
which an ultra cold atom incident on a possibly warm target will not
stick, even in the large $n$ limit where $n$ is the number of internal
degrees of freedom of the target.  Our treatment is non-perturbative
in which the full many-body problem is viewed as a scattering event
purely within the context of scattering theory.  The question of sticking is then
simply and naturally identified with the formation of a long lived
resonance.  One crucial physical insight that emerges is that the many
internal degrees of freedom serve to decohere the incident one body
wavefunction, thus upsetting the delicate interference process
necessary to form a resonance in the first place.  This is the
physical reason for not sticking. 
\end{abstract}

\section{Introduction}  

The problem of low energy sticking to surfaces has attracted much attention over the 
years\cite{lj,wal,Brenig,kohn,Bitt}.  The controversial question has been the ultralow
energy limit of the incoming species, for either warm or cold surfaces.  A battle has
ensued between two  countervailing effects, which we will call classical sticking and
quantum reflection. The concept of  quantum reflection is intimately tied into 
threshold laws, and was recognized in the 1930's by Lennard-Jones\cite{lj}.
Essentially, flux is reflected from a purely attractive potential  with a probability
which goes as $1-\alpha \sqrt{\epsilon}$, as $\epsilon\to 0$, where $\alpha$ is a constant and $\epsilon$ is the 
translational energy of the particle incident on the surface.  Classically the 
transmission probability is unity.   Reflection at long range prevents
inelastic processes from occurring, but if the incoming particle should 
penetrate into the strongly attractive region, the ensuing acceleration and 
hard collision with the repulsive short range part of the potential leads to a high probability of
inelastic processes and sticking.  

The blame for the quantum reflection can be laid at the feet of the WKB approximation,
which breaks down in the long range attractive part of the potential at 
low energy.  Very far out, the WKB is good even for low energy, because the potential is so nearly 
flat.  Close in, the kinetic energy is high, because of the attractive potential, even if the 
asymptotic energy is very low, and again WKB is accurate. But in between there is a breakdown, which has
been recognized and exploited by several groups\cite{jul,jul2,landau:lifshitz,joachain:qct,flam,cote}.  
We show  in the paper folling this one that  the
breakdown occurs in a region around
$\vert V \vert \approx \epsilon$; i.e. aproximately where the kinetic and potential energies are equal.

\begin{figure}
\centerline{\epsfig{figure=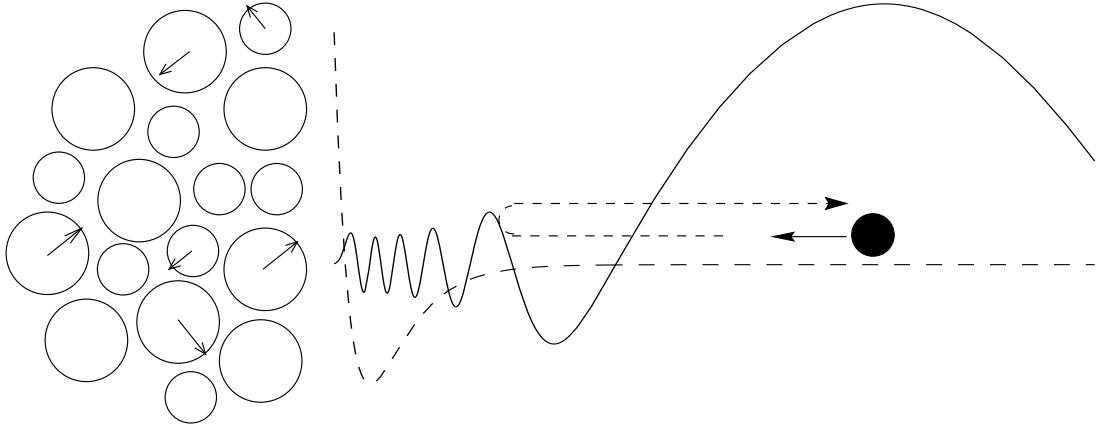,width=0.9\hsize}}
\hspace{0.1in}
\caption{The stationary state one body wavefunction of the
 incident atom moving in the $y$-independent mean potential felt by it.
  The amplitude inside the interaction region is supressed by 
$k_e \sim \sqrt{\epsilon}$.
 This is tantamount to the reflection of the atom.}
\label{fig:QR}
\end{figure}

\begin{figure}
\centerline{\epsfig{figure=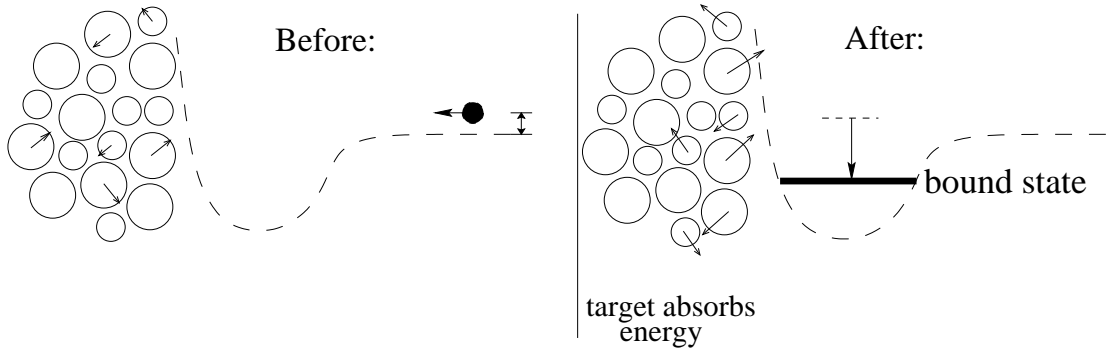,width=0.9\hsize}}
\hspace{0.1in}
\caption{A schematic view of a Feshbach resonance wherein the
 incident atom forms a long lived quasi-bound state with the
 target.  The many body wavefunction in this situation (not shown)
has a large amplitude in the `interior' region near the slab. }
\label{fig:Feshbach}
\end{figure}

It would seem that quantum reflection would settle the issues of sticking, since if the 
particle doesn't make it in close to the surface there is no sticking.
(Fig \ref{fig:QR})
There is one caveat, however,
which must be considered: quantum reflection can be defeated by the existence of a 
resonance in the internal region, i.e. a threshold resonance.
(Fig \ref{fig:Feshbach})

  The situation is very analogous to a 
high Q Fabry-Perot cavity, where  using  nearly 100\% reflective, parallel mirrors gives near 100\%
reflection
 except at very specific  wavelengths. At these specific energies  a resonace buildup occurs in the
interior of the cavity, permiting near 100\%  transmission. Such resonances are rare in a one dimensional
world, but the  huge number of degrees of freedom in a macroscopic solid particle makes resonance
ubiquitous. Indeed, the act of colliding with the surface, creating a phonon and 
dropping into a local bound state of the attractive potential describes a Feshbach resonance.
Thus, the resonances are just the sticking we are investigating, and we must not treat
them  lightly! Perhaps it is not obvious after all whether sticking occurs.

After the considerable burst of activity surrounding the sticking issue on the 
surface of liquid Helium\cite{doyle,hesuper},  and after a very well executed theoretical
study by Clougherty and Kohn\cite{kohn}, the controversy has settled down, and the 
common wisdom has grown that sticking does not occur at sufficiently 
low energy.  While we agree with this conclusion, we believe       the theoretical
foundation  for it is not complete, nor stated in a wide enough domain of physical
situations. For example, Ref.~\cite{kohn} treats only a harmonic slab with 
one or two
phonon excitation.  It is not clear  whether the results apply to a warm surface. On the
experimental side, even though quantum  reflection was observed from a liquid Helium
surface, that surface has a very low density of available states (essentially only the
ripplons) which could be a special case with respect to sticking. Thus, the need for
more rigorous and clear proof of non-sticking in general circumstances is evident.  This
paper gives such an analysis.  In a following paper, application is made  to specific
atom-surface and slab combinations, and the rollover to the sticking regime as energy is
increased (which can be treated essentially analytically) is given.

The strategy we use puts a very general and 
exact scattering formalism to work, providing a template into which to
insert the properties of our target and scatterer.  Then very general
results emerge, such as the non-sticking theorem at zero energy.  The
usual procedure of defining  model potentials and considering one
phonon processes etc. is not necessary. All such model potentials and
Hamiltonians wind
up as parameters in the R-matrix formalism.  The details of a particular
potential are of course important for quantitative results, but the range of
possible results can be much more easily examined by inserting various parameters into 
the R-matrix formalism. All the possible choices of R-matrix parameters give
the correct threshold laws.
Certain trends are built into the R-matrix formalism which are essentially independent
of the details of the potentials.

Before commencing with the R matrix treatment, we briefly consider the
problem perturbatively in order to better elucidate the role played by
quantum reflection.  We emphasize that none of the perturbation
section is actually necessary for our final conclusions.  

In a perturbative treatment for  our slab geometry, quantum reflection simply results in the
entrance channels' wave function (at threshold) having its amplitude
in the interaction region go to zero as $k_e\sim\sqrt{\epsilon}$ when
normalized to have a fixed incoming flux. ($k_e$ is the magnitude 
$|\vec k_e|$  of the incident wavevector of the incoming atom).
  The inelastic transition 
probabilities are proportional to
the potential weighted overlap of the channel wavefunctions and this
  immediately leads to the conclusion that the inelastic
probability itself vanishes as $k_e\sim\sqrt{\epsilon}$.  As mentioned,
this conclusion is shown to rigorously remain true using the R matrix.
We show in this paper that in spite of the inherently many-body nature of the
problem, in the ultra-cold limit  we can correctly
obtain the long-range form of the entrance channel's wavefunction by
solving for the one-dimensional motion in the 
long-range surface-atom attraction (i.e. the diagonal element of the
many-channel potential matrix).  This allows quantitative predictions of 
the sticking probability, which  we do in the following paper. There,
 we further exploit the perturbative point of
view together with an analysis of WKB to predict a `post-threshold'
behavior as quantum reflection abates, when the incoming energy is
increased.  

\section{Geometry and notation}

The incident atom is treated as a point particle at position $(x,y)$.
To keep the notation simple we leave out the $z$-coordinate and
confine our discussion to two spatial dimensions.  Thus a
cross-section will have dimensions of length etc.  It will be quite
obvious how and where $z$ may be inserted in all that follows.  
Let $u$ represent all the bound degrees of freedom of the
scattering target, which we take to be a slab of crystalline or amorphous
material.  Let $\Omega_c(u)$, $c = 1,2,\cdots$, be the manybody target wave functions in
the absence of interactions with the incident particle, and having
energy $E_c^{\rm target}$.  These are normalized as $\int_{\rm all\ u}
du\ {|\Omega_c(u)|^2} = 1$.  $x$ is the distance of the scatterer
(atom) from the face of the slab which is approximately (because the
wall is rough) along the line $x = 0$.  The internal constituents of
the slab lie to the left of $x = 0$ and the scatterer is incident from
the right with kinetic energy $\epsilon = \hbar^2 k_e^2 / 2m$.  The
total energy $E$ of the system is 
\begin{equation}
E = \epsilon + E^{\rm target}_e
\end{equation}
where $c = e$ is the index of the `entrance channel' i.e. the initial
internal state of the slab before the collision is $\Omega_e(u)$.  Notice
that we say nothing about the value of $E^{\rm target}_e$ itself.  In
particular the slab need not be cold.  $k_c$ is the magnitude of the wave vector $\vec k_c$ of the particle when it leaves the
target in the state $\Omega_c (u)$ after the collision.  Our
interest focusses on $k_e \rightarrow 0$. $k_e$
is the magnitude of the wavevector of the incoming particle. For the open
channels $c = 1,\cdots n$ (this defines n) for which $E > E^{\rm target}_c$ 
\begin{equation}
    k_c \equiv \sqrt{{2m(E - E_c^{\rm target})\over \hbar^2} }
    \qquad (c \leq n)
    \; ;
\end{equation}
whereas for the closed channels ($c > n$), $E < E_c^{\rm target}$ and
\begin{equation}
    k_c\equiv i\sqrt{{2m (E_c^{\rm target} - E)\over \hbar^2}} \equiv i\kappa_c
    \qquad (c > n)
    \; .
\end{equation}
$\kappa_c > 0$. 
We will use $(k_{c x}, k_{c y})$ as the $x,y$ components of $\vec
k_c$.  Let $U_{\rm int} (x,y,u) = (2m / \hbar^2) V_{\rm int}(x,y,u)$,
where $V_{\rm int}(x,y,u)$ describes quite generally the interaction
potential between the incident atom and all the internal degrees of
freedom of the slab.  For simplicity we assume for the moment that
there is no interaction between slab and atom for $x > a$.

\section{Preliminaries: Perturbation}
\label{Perturbation}

As stated above, we excercise the perturbative treatment for insight only; our final
conclusions are based on   nonperturbative arguments.

We treat the interaction $U_{\rm int}(x,y,u)$ between slab and atom by 
separating out a `mean' potential felt by the atom that is independent
of $y$ and $u$; call it $U^{(0)}(x)$.  The remainder
$U^{(1)}(x,y,u)\equiv U_{\rm int}(x,y,u) - U^{(0)}(x)$ is treated as a perturbation.

Now the incident beam is scattered by the entire length (say from $y = -L \ {\rm to}\  L
= 2L$) of wall which it illuminates.  If all measurements are made close to the wall so that
its length $2L$ is the largest scale in the problem, then it is
appropriate to speak of a cross-section per unit length of wall, a
dimensionless probability.  More specifically, we will assume that the
matrix elements $U^{(1)}_{cc'}(x,y) \equiv \int\limits_{\rm all \ u} du \, \Omega_c^*(u)
    U^{(1)}(x,y,u) \Omega_{c'}(u)$ of the perturbation $U^{(1)}(x,y,u)$ in the
    $\Omega_c(u)$ basis are given by the simple form
    $U^{(1)}_{cc'}(x,y) = U^{(1)}_{cc'}(x)f(y)$ for $y \in [-L, L]$
    and 0 elsewhere.  $f(y)$ is a random persistent (does not
    die to 0 as $|L|\rightarrow \infty$) function that models the random
    roughness of the slab and  is characterized by its so-called spectral density
    function $S$, a smooth positive-valued non-random function, such
    that 
\begin{equation}
\left|\int\limits_{-L}^L dy \,e^{i k y} f(y)\right|^2 \equiv 2L
S(k)\hspace{0.5cm}\forall k
\end{equation}  
as $L\rightarrow \infty$.

Now, applying either time-independent perturbation (equivalently the
Born approximation for this geometry) or time-dependent perturbation
theory via the Golden Rule, gives that the cross-section per unit
length of wall for inelastic scattering to a final channel $c$ is 
\begin{equation}
P^{\rm in}_{c\leftarrow e} (\theta) = {2\pi \over k_e} 
\left(\int\limits_{-\infty}^a dx'
\phi(x';k_{cx})
U^{(1)}_{ce}(x')\phi(x';k_{ex})\right)^2
S(k_{cy} - k_{ey})
\label{P_in}
\end{equation}
where $\phi(x;k_x)$ is the solution of the o.d.e.
\begin{equation}
\left({d^2 \over dx^2} - U^{(0)}(x) + k_{x}^{2}\right)\phi(x;k_x) = 0
\end{equation}
which is regular or goes to zero as $x\rightarrow - \infty$ inside the
slab and is normalized as
\begin{equation}
\phi(x;k_x) \sim \sin (k_x x+\delta)\hspace{0.5 cm} {\rm as} x\rightarrow
\infty
\label{normalization}
\end{equation}
Accepting for the moment that as $k_e \rightarrow 0$ the amplitude of $\phi (x;k_{ex})$
in the internal region $x < a$ goes to zero as $k_e\sim \sqrt
{\epsilon}$, 
then the square of the overlap integral in Eq.~(\ref{P_in}) behaves as
$k_e^2$,
because by our proposition the amplitude of 
$\phi(x';k_{ex}) \sim k_{ex} \sim k_e$.
Together with the $1/k_e$ prefactor we get an overall behavior of
$k_e$ for the inelastic probability as claimed.  \newline
To show  that  indeed as $k_e \rightarrow 0$ the amplitude of $\phi (x;k_{ex})$
in the internal region $x < a$ goes to zero as $k_e\sim \sqrt
{\epsilon}$, 
 we temporarily disregard  the required normalization of
$\phi(x;k_x)$ of Eq.~(\ref{normalization}) and fix its initial conditions (slope and
value) at some point inside the interaction region $x<a$ such that the
regularity condition is ensured.  We then integrate out to $x=a$. Let
us denote this unnormalized solution with a prime, as $\phi'(x;k_x)$.
The point is   for $k_x$ varying near 0, both $v$(the value)
and $s$(the slope) that the solution emerges with at $x=a$, are
independent of $k_x$ and in fact the interior solution thus obtained
is itself independent of $k_x$. This is   because the
local wave vector $k(x) = \sqrt{2m(\epsilon - U(x))/\hbar^2}$
essentially stays the same function of $x$ for all $\epsilon$ near 0.
Therefore for $x>a$ $\phi(x;k_x)$ 
continues onto 
\begin{equation}
 v \cos [k_x(x-a)] + {s \over k_x} \sin[k_x(x-a)] \hspace {.5 cm} x>a
\end{equation}
This is a phase-shifted sine wave of amplitude $\sim 1/k_x$.  We
must enforce the normalization of Eq.~(\ref{normalization}) and get 
$\phi(x;k_x) \sim k_x \phi'(x;k_x)$. As a
result, the interior solution gets multiplied by $k_x$ and we thereby
have our result. 
$\phi(x;k_x) $ is the solution of a one-dimensional Schrodinger
equation for the incoming particle in the one-dimensional long-range
potential created by the slab. The suppression of its amplitude
by $\sqrt{\epsilon}$ near the slab is due to the reflection
it suffers where the interaction turns on. Within the perturbative
set-up the non-sticking conclusion is then already foregone\cite{lj}.  

The problem is whether we can really accept this verdict of the 
one-dimensional unperturbed solution, when in fact we know that the
turning on of the perturbation (many body interactions) causes a
multitude of resonances to be created, internal resonances being 
exactly the situation in which the Proposition above is known to badly
fail. It appears that the perturbation is in no sense a small physical
effect. 
Therefore a nonpeturbative 
approach is needed.  Here we use R-matrix theory in its general form to 
accomplish the task.

\section{$\bS$-matrix and $\bR$-matrix}
\label{s_matrix_and_r_matrix}
One point that the preceding section has made clear is that it is the 
energies (both
initial and final) in the $x$-direction, perpendicular to the slab
that are most relevant.  In fact as regards the final form of our
answers the motion of the $y$ degree of freedom may as well have been
the motion of another internal degree of freedom of the slab.  In
other words, mathematically speaking, the $y$ degree of freedom may be
subsumed by incorporating it as just another $u$.  For example, we may
imagine the incident atom being confined in the $y$-direction by the
walls of a wave-guide at $y = -L {\rm and} L$ that is large enough so
that it could not possibly change the physics of sticking.  Then we
quite rigorously have a bound internal state of the form 
\begin{equation}
\Omega_{c,n} (y,u) = \Omega_c (u) \sin {n\pi y \over L}
\label{extended_internal_state}
\end{equation}
$x$ is now the only scattering degree of freedom.  There will be no
necessity in carrying along the extra index $n$ and variable $y$ as in
Eq.~(\ref{extended_internal_state}), and we will simply continue to write
$\Omega_c (u)$ instead.  Thus with this understanding, the problem is
essentially one-dimensional in the scattering degree of freedom.  

We  proceed to derive the expression for the $\bS$ matrix in terms of the
so-called $\bR$ matrix, and derive the structure of the $\bR$ matrix.
For simplicity we continue to assume for the moment
that there is no interaction for $x > a$.  Then for $x > a$, the
scattering wavefunction of the interacting system corresponding to the
scattering particle coming in on one entrance channel, say $c = e$,
with energy $\epsilon = \hbar^2 k_e^2/(2m)$ is
\begin{equation}
    \psi(x,u) = \sum_{c=1}^\infty 
    \left( {e^{-i k_e x}\over \sqrt{k_e}} \delta_{ce} 
        - {e^{i k_c x}\over \sqrt{k_c}} S_{ce} \right)
    \Omega_c(u) \qquad x > a
    \label{asymptotic_wavefunction}
\end{equation}
where the sum must include all channels, even though the open channels
are finite in number. 
The factors of $k_c^{-1/2}$ in Eq.~(\ref{asymptotic_wavefunction})
mean that the flux in each channel is proportional only to the square
of the coefficient and hence ensure the unitarity of $\bS$.  With this
convention, the open-open part of the $\bS$-matrix---the $n \times n$
submatrix $S_{cc'}$ with $c, c' = 1, 2, \ldots, n$---is unitary.
$\sqrt{k_c} \equiv e^{i \pi/4} \sqrt{\kappa_c}$ may be arbitrarily
chosen since it cannot affect the open-open part of $\bS$.

$\bS$ is found in analogy to the one-dimensional case by introducing
the matrix version of the inverse logarithmic derivative at $x=a$
called $\bR (E)$ the Wigner $\bR$-matrix defined by
\begin{equation}
\vec{v}=\bR (E)\  \vec{s} 
\label{vequalRs}
\end{equation}
where the components of $\vec{v}$ and $\vec{s}$ are the expansion
coefficients of $\psi(x=a,u)$ and ${\partial\psi (x=a,u)\over \partial
  x}$ respectively in the $\Omega_c(u)$ basis. Supposing
${\partial\psi (x=a,u)\over \partial x}$ to be known, we will (like in
electrostatics) use the Neumann Green's function $G_N(x,u;x',u')$ to
construct $\psi(x,u)$ everywhere in the interior $x<a$. $\psi(x,u)$
satisfies the full Schr\"{o}dinger equation with energy $E$. We need
$\chi_\lambda(x,u) \ \lambda=1,2,\cdots,$ the normalized
eigenfunctions of the full Schr\"{o}dinger equation in the interior
$x<a$ with energies $E_\lambda$, satisfying Neumann boundary
conditions ${\partial\chi (x=a,u)\over \partial x} = 0$. So
\begin{eqnarray}
&&\left( {-\hbar^2 \over 2m} \nabla^2 + V_{\rm int}(x,u) - E \right)
    \psi(x,u) = 0 \\
&&\left( {-\hbar^2 \over 2m} \nabla^2 + V_{\rm int}(x,u)
    - E_\lambda \right) \chi_\lambda(x,u) = 0 \\
&&\left( {-\hbar^2 \over 2m} \nabla^2 + V_{\rm int}(x,u) - E \right)
    G_N(x,u;x',u') = \delta(x-x') \delta(u-u') 
\end{eqnarray}
where $\nabla^2 \equiv {\partial^2 \over \partial x^2} + {\partial^2
  \over \partial u^2}$ and
\begin{eqnarray}
{\partial G_N (x=a,u;x',u') \over \partial x} &=& 0 
\hspace{1cm} {\mbox and} \hspace{1cm} 
{\partial \chi(x=a,u) \over \partial x} = 0 \\
\Rightarrow G_N(x,u;x',u') &=& \sum\limits_{\lambda=1}^\infty
    {\chi_\lambda(x,u)\chi_\lambda(x',u') \over E_\lambda - E}
\end{eqnarray}
$G_N$ is symmetric in the primed and unprimed variables.  By Stokes'
Theorem,
\begin{equation}
    (-\hbar^2/2m) \int\limits_{x'<a}dx'\int\limits_{\text{all
    u'}}du'
    \left( \phi_{1} \nabla'^{2}\phi_{2}
        - \phi_{2} \nabla'^{2}\phi_{1} \right)
    = (-\hbar^2/2m) \int\limits_{\text{x'=a, all u'}}du'
    \left( \phi_{1} \nabla_{\hat n}'\phi_{2}
        - \phi_{2} \nabla_{\hat n}'\phi_{1} \right)
\end{equation}
where $\nabla_{\hat n}'(\cdot) \equiv {\hat x}'(\cdot)  \cdot \nabla'$ with
$\phi_1 = \psi(x',u')$ and $ \phi_2 = G_N(x,u;x',u')$ gives
\begin{equation}
\psi(x,u) = {\hbar^2 \over 2m} \int\limits_{\rm all \ u'} du'
\ G_N(x,u;x',u')  {\partial \psi (x'=a,u') \over \partial x'} \hspace{1cm}
x<a
\label{neumann_psi}
\end{equation} 
Put $x=a$ and it is deduced using Eqs.~(\ref{vequalRs}) and
(\ref{neumann_psi}) together that
\begin{equation}
R_{cc'}(E) =  \sum_{\lambda = 1}^{\infty}
        \frac{\gamma_{\lambda c}\gamma_{\lambda c'}}{E_{\lambda} - E}
\label{R_matrix_expansion}
\end{equation}
where $\gamma_{\lambda c}= \sqrt{{\hbar^2 \over 2m}} \int\limits_{\rm
  all \ u} du\ \chi_\lambda(a,u) \Omega_c(u)$.

\subsection{The $\bS$ matrix}
\label{subsec:S_matrix}

Now shifting attention to the outside ($x>a$), we see that we can
compute both $\nabla_{\hat n}\psi(a,u)$ and $\psi(a,u)$ on the surface
$x = a$ using the asymptotic form of Eq.~(\ref
{asymptotic_wavefunction}) which automatically gives these expanded in
the $\Omega_{c}(u)$ basis.  Writing the matrix Eq.~(\ref{vequalRs}) is
now simple.  It is best to do it all in matrix notation, and thus be
able to treat all possible independent asymptotic boundary conditions
simultaneously.

Let $e^{i k x}$, $\sqrt{k}$ and $1/\sqrt{k}$ be diagonal matrices with
diagonal elements $e^{i k_{c} x}$, $\sqrt{k_{c}}$ and
$1/\sqrt{k_{c}}$.  Then Eq.~(\ref{vequalRs}) reads
\begin{equation}
    {e^{-i k a}\over\sqrt{k}} - {e^{i k a} \over \sqrt{k}} \bS
    = i \bR k
    \left( {-e^{-i k a}\over\sqrt{k}} - {e^{i k a}\over\sqrt{k}} \bS \right)
    \label{equation_for_S}
    \; .
\end{equation}
Each column $c=1,\ldots,n$ of the matrix equation above is just
Eq.~(\ref{vequalRs}) for the solution corresponding to an incoming
wave only in channel $c$ (For $c > n$ the wavefunctions blow up as $x
\rightarrow \infty$). Remembering that non-diagonal matrices don't
commute, we solve for $\bS$ to get
\begin{equation}
    \bS=e^{-i k a} \sqrt{k} {1\over 1-i \bR k} (1 + i \bR k) {1\over
      \sqrt{k}}e^{-i k a}
    \label{S_matrix_1}
\end{equation}
or, with some simple matrix manipulation,
\begin{equation}
    \bS=e^{-i k a}{1\over 1-i\sqrt{k} \bR \sqrt{k}} (1 + i\sqrt{k} \bR
    \sqrt{k})e^{-i k a}
    \label{S_matrix_2}
    \; .
\end{equation}

\section{$\bS$ matrix near a resonance}
\label{s_matrix_near_resonance}
As discussed in the introduction, the resonances are a key to the sticking issue.
Sticking is essentially a long lived Feshbach resonance in which energy has been supplied to 
surface and bulk degrees of freedom, temporarily dropping the scattering particle into 
a bound state of the attractive potential.  Thus we must study resonances in 
various circumstances in the low incident translational energy regime.
We  derive the approximation for $\bS(E)$ near $E = E_0$, a resonant energy of
the compound system.     $E_0$ is the total energy of the joined (resonant)
system.
Within the R-matrix approach,  the $\chi_\lambda (x,u)$ of section
\ref{s_matrix_and_r_matrix} are  bound, compound states
 with Neuman boundary conditions at $x = a$.  $R$-matrix theory 
properly couples these bound state to the continuum, but some
of the eigenstates are nonetheless weakly coupled to the continuum, as evidenced  
by small values of the 
$\gamma_{\lambda c}$'s of section \ref{s_matrix_and_r_matrix}; these  are the
measure of the strength of the continuum couplings.    While every one of the
$R-$matrix bound
states will result in a pole $E_\lambda$ in the $R$ matrix expansion, only the
weakly coupled ones are the true long lived Feshbach resonances of physical
interest.  It is also helpful to know that the values of these
`truly' resonant poles at $E_\lambda$ are the most stable to changes
in the position $x = a$ of the box.  This in fact provides one
unambiguous way to identify them.  
Our purpose here is to derive the resonant approximation to
the $\bS$ matrix in the vicinity of one of these Feshbach resonances.
We do so using the form of the $\bR$-matrix in
Eq.~(\ref{R_matrix_expansion}).  Note that the
energy density $\rho(E) = 1 / D(E)$  of these Feshbach resonances will be large
because of the large number of degrees of freedom of the target.  $D(E)$ is the
level spacing of the  quasibound, resonant  states.
\subsection{Isolated Resonance}
\label{subsec:isolated resonance}
As mentioned, the point of view we will take is to identify a resonant energy with a
particular pole $E_\lambda$ in the $\bR$ matrix expansion of
Eq.~(\ref{R_matrix_expansion}).  Those $E_\lambda$
corresponding to resonances are a subsequence of the $E_\lambda$
appearing in the expansion in Eq.~(\ref{R_matrix_expansion}).
For $E$ near a well isolated resonance at $E_{\lambda}$
we separate the sum-over-poles expansion of the R-matrix into a single
matrix term  having elements ${\gamma_{\lambda c} \gamma_{\lambda c'} \over E_\lambda - E}$,
plus a sum over all the remaining terms, call it $N$. If the energy
interval between $E_{\lambda}$ and all the other poles is
large compared to the open-open residue at $E_{\lambda}$ then  we may expect
that the
$ n\times n$ open-open block of $N$ will have all its elements to be
small. 
 Then rewriting the  inverse in Eq.~(\ref{S_matrix_2}) 
\begin{equation}
{1 \over1- i \sqrt{k} \bR \sqrt{k}} \equiv 
{1 \over 1-i \left(M + {V \over E_{\lambda}-E} \right)}
\end{equation}
where $M\equiv \sqrt{k} N \sqrt{k}$ and
 $V_{c c'}\equiv 
(\sqrt{k_c} \gamma_{\lambda c})
(\sqrt{k_{c'}} \gamma_{\lambda c'})$, and setting $M=0$ allows us to simplify the central
term in Eq.~(\ref{S_matrix_2}) exactly. (We will return to the case $M\not=$0.)
\begin{eqnarray}
    &&{1 \over 1 - i \sqrt{k}\bR \sqrt{k}} (1 + i \sqrt{k}\bR \sqrt{k}) \\
    &=& 1 + {1 \over 1 - i \sqrt{k}\bR \sqrt{k} } 2 i \sqrt{k} \bR \sqrt{k} \\
    &=& 1 + {1 \over 1 - {i V  \over E_\lambda - E}}2 i {V \over
      E_\lambda - E} \hspace{.5cm} ({\rm with} M=0)\\
    &=& 1 + {1 \over E_\lambda - E - i V } 2 i V \\
    &=& 1 + {1 \over E_\lambda - E - i(\Gamma_{\lambda}/2+i\Delta E)} 2 i V k
\end{eqnarray}
where we used 
\begin{eqnarray}
    V^2&=& \left( (\gamma_{\lambda 1}^2 k_1 + \cdots +
    \gamma_{\lambda n}^2
    k_n) + (\gamma_{\lambda(n+1)}^2 \kappa_{n+1} + \cdots) \right)V\\
    &\equiv& \left(\left({\Gamma_{\lambda 1} \over 2} +  \cdots +
    {\Gamma_{\lambda n}
          \over 2} \right) + i(\gamma_{\lambda(n+1)}^2 \kappa_{n+1} +
    \cdots) \right)V
\\
    &\equiv& \left({\Gamma_\lambda \over 2} + i \Delta E_\lambda\right)V
\end{eqnarray}
to get the identities
\begin{eqnarray}
    [E_\lambda - E - i V] V &=& [E_\lambda - E - i
    (\Gamma_\lambda/2+i\Delta E)]V \\
    \Rightarrow {1 \over E_\lambda - E - i(\Gamma_\lambda/2+i\Delta E) } V &=& {1 \over
      E_\lambda - E - i V}V 
\end{eqnarray}
Also define $(\Gamma_{\lambda c}/2)^{1/2} \equiv \gamma_{\lambda c}
\sqrt{k_c}, c = 1, 2, \cdots ,n$. This defines the sign of the
square-root on the lhs. to be the sign of $\gamma_{\lambda c}$ and
allows the convenience of expressing things in terms of the
$\Gamma_{\lambda c}$'s and their square-roots, and not having to use 
the $\gamma_{\lambda c}$'s themselves. Thus we arrive at 
\begin{equation}
    S_{cc'}=e^{-i k_{c} a} \left(\delta_{c c'} + {i \Gamma_{\lambda c}^{1/2}
          \Gamma_{\lambda c'}^{1/2}
          \over E_\lambda^{(r)} - E - i \Gamma_\lambda / 2}\right) 
          e^{-i k_{c'} a}
\label{S_matrix_form}
\end{equation}
where $E_\lambda^{(r)} \equiv E_\lambda + \Delta E_\lambda$, for the 
$n\times n$ open-open unitary block of $S$ in the neighbourhood of a
single isolated resonance after neglecting the contribution of the
background matrix $M$.
For us the
essential point is that
\begin{equation}
    \Gamma_{\lambda c} = 2\,k_c(E)\gamma_{\lambda c}^2 ,
\label{E_dependence_of_Gamma}
\end{equation}
that the partial widths $\Gamma_{\lambda c}$ depend on the energy $E$,
through the kinematic factor $k_c(E)$. 
Mostly this energy dependence is small and irrelevant except 
where the $k_c$'s and hence 
$\Gamma_{\lambda c}$'s are varying near 0.
These are the partial widths of the open channels near threshold.
Hence $|S_{c e}|^2$ ($c\not=e$) an inelastic probability
behaves like 
$k_e \sim \sqrt{\epsilon}$ when the entrance channel is at threshold.
Including the background term ($M\not=$ 0) does not change this. 
To see this we may perform the inverse in Eq.~(\ref{S_matrix_2})
to first order in $M$ and then get an additional contribution 
of the terms
\begin{equation}
e^{-i k a} \left(
{2 i \over 1-{i V\over E_\lambda-E}} M +
{1 \over 1-{i V\over E_\lambda-E}} +
{1 \over 1-{i V\over E_\lambda-E}} i M 
{1 \over 1-{i V\over E_\lambda-E}} 2 i V
\right) e^{-i k a}
\label{background_addition}
\end{equation}
to the $S$-matrix.
Now, both $M$ and $V$ have a factor of $\sqrt{k_c}$ multiplying their
$c$th columns (and rows) from their definitions and so a matrix
element $b_{c c'}$ of the matrix in parentheses in 
Eq.~(\ref{background_addition}) will have a $\sqrt{k_c}$ and
$\sqrt{k_{c'}}$ dependence.
An inelastic  element of $S(c\not=c')$ would now take the form 
\begin{equation}
S_{cc'}=e^{-i k_{c} a} \left(b_{c c'} + {i \Gamma_{\lambda c}^{1/2}
          \Gamma_{\lambda c'}^{1/2}
          \over E_\lambda^{(r)} - E - i \Gamma_\lambda / 2}\right) 
          e^{-i k_{c'} a}
\label{S_matrix_form_with_background},
\end{equation}
As mentioned our interest is in the case  when the entrance channel is 
at threshold so that this
dependence is $\sqrt{k_e}$, making the inelastic probability
$|S_{ce}|^2$ still continue to behave as 
$k_e\sim \sqrt{\epsilon}$.

\subsection{Overlapping Resonances}
\label{overlapping_resonances}
Here we require the form of the $\bS$ matrix near an energy $E$ where many
of the quasibound states may be simultaneously excited, i.e. the
resonances overlap.   Again, neglecting background
for the moment, the $\bS$ matrix is simply taken to be a sum over the
various resonances.
\begin{equation}
    \bS=1- \sum\limits_{\lambda} {i A_{\lambda} \over E - E_\lambda^{(r)} + i
      \Gamma_\lambda/ 2}
\label{resonant_S_matrix}
\end{equation}
where $A_\lambda$ is a $n\times n$ rank 1 matrix with the $c c'$th
component as $\Gamma_{\lambda c}^{1/2} \Gamma_{\lambda c'}^{1/2}$.
There is no entirely direct justification of this form, but one can
see that there is much which it gets correct.  

The $A_{\lambda}$ are symmetric, hence $\bS$ is symmetric.  Obviously
it has the poles in the right places allowing the existence of
decaying states with a purely outgoing wave at the resonant energies.
A crucial additional assumption that also makes $\bS$ approximately
unitary is that the signs of the $\Gamma_{\lambda c}^{1/2}$ are random
and uncorrelated both in the index $\lambda$ as well as $c$,
regardless of how close the energy intervals involved may be.  One
simple consequence is that we approximately have that
\begin{equation}
A_{\lambda}A_{\lambda'} = \delta_{\lambda
  \lambda'}\Gamma_{\lambda}A_{\lambda}
\label{A_orthogonality}
\end{equation}
in the sense that the l.h.s. is negligible
for $\lambda \not= \lambda'$ in comparison to the value for $\lambda =
\lambda'$.  With Eq.~(\ref{A_orthogonality}) it is easy to verify the
approximate unitarity of $\bS$.  

We investigate now the onset of the overlapping regime as $E$ increases.
$D(E)$, the
level spacing of the resonant $E_\lambda^{(r)}$, is a rapidly decreasing function of
its argument.  On the other hand, $\Gamma_\lambda = \Gamma_{\lambda 1}
+ \Gamma_{\lambda 2} + \cdots + \Gamma_{\lambda n}$, and since more
channels are open at higher energy, $\Gamma_\lambda$ is increasing
with the energy of the resonance.  The widths must therefore
eventually overlap, and $\Gamma_\lambda \gg D\left( E_\lambda^{(r)}
\right)$ for the larger members of the sequence of
$E_\lambda^{(r)}$'s.  In this regard there is a useful estimate due to
Bohr and Wheeler  \cite {bohr_wheeler}, that for $n$ large
\begin{equation}
  {\Gamma_\lambda \over D(E_\lambda^{(r)})} \simeq n \ .
\label{}
\end{equation}
Appendix \ref{Bohr_Wheeler} derives this using a phase space argument.
Here we point out that this is entirely consistent with the assumption
of the random signs, indeed requiring it to be true.  Take for example
a typical inelastic amplitude
\begin{equation}
\bS_{c c'} = -i\sum\limits_{\lambda}{ \Gamma_{\lambda c}^{1/2}
          \Gamma_{\lambda c'}^{1/2}
          \over E_\lambda^{(r)} - E - i \Gamma_\lambda / 2}
\hspace{.7cm} (c\not=c')
\label{}
\end{equation}
First let us note that the $\Gamma_{\lambda}$ being the sum of many
random variables (the partial widths $\Gamma_{\lambda c}$) do not
fluctuate
 much.  Let $\Gamma$ denote their
typical value over the $n$ overlapping resonances.  
Also since $\Gamma = nD$ it follows that the typical
size of a partial width $\Gamma_{\lambda c}$ is $D$.  Therefore the typical size of
the product $ \Gamma_{\lambda c}^{1/2}\Gamma_{\lambda c'}^{1/2}$ is $D$ but
these random variables fluctuate randomly over the index $\lambda$,
and moreover
the sign is random.  Thus for energies in the overlapping  domain $S_{c c'}$
is a sum of $n$ complex numbers each of typical size
$D/\Gamma=1/n$, but random in sign.  This makes for a sum of order $1/\sqrt{n}$.  Clearly this
is as required to make the $n \times n$ matrix $\bS$ unitary.  Note
that the above argument fails (as is should) if $c\not=c'$
because then the signs of $ \Gamma_{\lambda
  c}^{1/2}\Gamma_{\lambda c}^{1/2} = \Gamma_{\lambda} > 0$ are of
course not random.  

Unlike the case of the isolated resonance, the S-matrix
elements here are smoothly varying in E. Addition of a background term  
$B_{c c'}$
\begin{equation}
\bS_{c c'} = B_{c c'} -i\sum\limits_{\lambda}{ \Gamma_{\lambda c}^{1/2}
          \Gamma_{\lambda c'}^{1/2}
          \over E_\lambda^{(r)} - E - i \Gamma_\lambda / 2}.
\label{}
\end{equation}
just shifts this smooth variation by a constant.  If $B_{c c'}$ is
also thought of as arising from a sum over the individual backgrounds
then for the same reasons as discussed at the end of the preceding
section $|B_{c e}|^2 \sim k_e \sim \sqrt{\epsilon}$ for an
entrance channel near threshold.
For simplicity we will continue to take $B_{c c'}$ to be 0 and look at
the case with background in the appendix.

\section{$\bQ$-matrix and Sticking}

 From the viewpoint of scattering theory, the sticking of the incident
particle to the target is just a long-lived resonance.  It is natural
then to investigate the time-delay for the collision.  Smith\cite{smith}
introduced the collision lifetime or $Q$-matrix
\begin{equation}
   \bQ \equiv i\hbar \bS {\partial \bS^\dagger \over \partial E}
\label{Q_matrix_definition}
\end{equation}
which encapsulates such information.  We review some of the relevant
properties of $\bQ$.  The rhs of Eq.~(\ref{Q_matrix_definition})
involves the `open-open' upper left block of $\bS$ so that $\bQ$ is
also an $n \times n$ energy-dependent matrix, having dimensions of
time.  For 1-dimensional elastic potential scattering $\bS = e^{i
  \phi(e)}$ and $\bQ$ reduces to the familiar time delay
$i\hbar{\partial \phi(E) \over \partial E}$.  If $\vec v$ is a vector
whose entries are the coefficients of the incoming wave in each
channel then ${\vec v}^{\rm t r} \bQ(E) \vec v$ is the average delay
time experienced by such an incoming wave.  Because physically the
particle is incident on only one channel, $\vec v$ consists of all 0's
except for a 1 in the $e$th slot so that the relevant quantity is just
the matrix element $\bQ_{ee}(E)$.  Smith shows that this delay time is
the surplus probability of being in a neighborhood of the target
(measured relative to the probability if no target were present)
divided by the flux arriving in channel $e$.  This matches our
intuition that when the delay time is long, there is a higher
probability that the particle will be found near the target.

Furthermore, as a Hermitian matrix, $\bQ(E)$, can be resolved into its
eigenstates $\vec v^{(1)} \cdots \vec v^{(n)}$ with eigenvalues $q_1
\cdots q_n$.  The components of $\vec v^{(1)}$ are the incoming
coefficients of a quasi-bound state with lifetime $q_1$ and so on.
Then
\begin{equation}
{\vec v}^{\rm t r} \bQ(E) \vec v = \sum\limits_{j = 1}^n q_j |\vec v^{(j)} \cdot \vec v |^2 .
\label{i_transpose_Q_i}
\end{equation}
As can be seen from this expression, the average time delay results,
in general, from the excitation of multiple quasi-stuck states each
with its lifetime $q_j$ and probability of formation $|\vec v^{(j)}
\cdot \vec v |^2$.  However, we will find that using our resonant
approximation to the $\bS$ matrix near a resonant energy
$E_\lambda^{(r)}$ the time delay will consist of only one term from
the sum on the rhs of Eq.~(\ref{i_transpose_Q_i}), all the other
eigenvalues being identically 0.

Using equation Eq.~(\ref{Q_matrix_definition}), 

\begin{eqnarray}
    \bQ(E) = i \hbar \Biggl( \sum\limits_{\lambda'} {-i A_{\lambda'} \over
      \left [E- E_{\lambda'}^{(r)} - i \Gamma_{\lambda'}/2 \right] ^2} -\sum
    \limits_{\lambda \lambda'} {A_{\lambda} A_{\lambda'} \over
      \left[E - E_\lambda^{(r)} + i \Gamma_\lambda / 2 \right ] \left [E-
          E_{\lambda'}^{(r)} - i \Gamma_{\lambda'}/2 \right] ^2} \Biggr) 
\label{}
\end{eqnarray}
which using Eq.~(\ref{A_orthogonality}) simplifies to 
\begin{equation}
    = \sum\limits_{\lambda} {\hbar \over (E- E_\lambda^{(r)})^2 +
      (\Gamma_\lambda/2)^2} A_\lambda \ ,
\end{equation}
a remarkably simple answer.  We need $Q_{e e}(E)$, where $e$ is the
entrance channel.

\begin{eqnarray}
    Q_{e e}(E) &=& \sum\limits_{\lambda} {\hbar \Gamma_{\lambda e} \over (E-
      E_\lambda^{(r)})^2 + (\Gamma_\lambda/2)^2} \\
    &=& \sum\limits_{\lambda} \left( {\hbar \Gamma_{\lambda} \over (E-
          E_\lambda^{(r)})^2 + (\Gamma_\lambda/2)^2} \times {\Gamma_{\lambda e}
          \over \Gamma_\lambda} \right) 
\label{Q_ee}
\end{eqnarray}
where the second equation has the interpretation (for each term) as
the life-time of the mode, multiplied by the probability of its
formation.  Note how for each resonance $E_\lambda^{(r)}$ there is
only one term corresponding to the decomposition of
Eq.~(\ref{i_transpose_Q_i}). The actual measured lifetime is the
average of $Q_{e e}(E)$ averaged over the energy spectrum $|g(E)|^2$
of the collision process.
 
\subsection{Energy averaging over spectrum}
With the target in state $\Omega_{e}(u)$ where $c=e$ is the entrance
channel, the energy of the target is fixed, and the time-dependent
solution will look like
\begin{equation}
    \psi(x,u,t) = \int dE
    \left( g(E) \sum\limits_{c=1}^{\infty} \left({e^{-i k_{c}(E) x}
    \over \sqrt{k_e(E)}}
      \delta_{c e} - {e^{i k_{c}(E) x}\over \sqrt{k_c(E)}} S(E)_{c e} \right) 
    \Omega_{c}(u) \right). 
\end{equation}
Recall, $E$ is the total energy of the system.  We are interested in the threshold situation where the incident
kinetic energy of the incoming particle $\epsilon \rightarrow 0$.
This can be arranged if $g(E)$ is peaked at $E_0$ with a spread
$\Delta E$ such that i) $E_0$ is barely above $E_e^{\rm target}$ and
ii) $\Delta E=\delta \epsilon$ is some small fraction of $\epsilon$, the mean energy
of the incoming particle.  The second condition ensures that we may
speak unambiguously of the incoming particle's mean energy.  So,
\begin{eqnarray} 
  \langle Q_{e e}(E)\rangle &\equiv& \int dE |g(E)|^2 Q_{e e}(E) \\
                            &\simeq& {1 \over \Delta E} \int dE Q_{e e}(E)
\label{mean_Q}
\end{eqnarray} 
$\langle \rangle$ denotes the average over the $\Delta E$ interval.  
Now $Q_{e e}(E)$ is just a sum of Lorentzians centred at the
$E_\lambda^{(r)}$'s with width $\Gamma_\lambda$ and Eq.~(\ref{mean_Q})
is just a measure of their mean value over the $\Delta E$ interval.

So long as the $\Delta E$ interval around which we are averaging,
is broad enough to straddle many of these Lorentzians, the mean height is just
\begin{equation}
\label{halfway}
    {1 \over \Delta E}\times \rho(E) \Delta E \times 
    {\hbar \pi \Gamma_{\lambda e} \over \Gamma_\lambda}
\label{mean_Q_explicit}
\end{equation}
where the second factor is the number of Lorentzians in the $\Delta$E
interval and the third factor is the area under the `$\lambda$th'
Lorentzian.  This is true regardless of whether or not they are
overlapping. 
It will be convenient to write
$\Gamma_\lambda$ as
\begin{equation}
 \Gamma_\lambda =  n\,\times 2 \bar k_\lambda\, {\rm var}(\gamma_\lambda)
\label{Gamma_lambda_taken_apart}
\end{equation}
where ${\rm var}(\gamma_\lambda)$ is the variance of the set of 
${\gamma_{\lambda c}}'s $
over the $n$ open channels and $\bar k_\lambda$ is a mean or effective
wavenumber $k_c$ over the open channels, which for a particular
realization $\lambda$ we take to be defined by
Eq.~(\ref{Gamma_lambda_taken_apart}) itself.  Let $\langle\ \rangle$
denote the average over the occurrences of the quantity in the $\Delta
E$ interval.  $\Gamma \equiv \langle\Gamma_\lambda\rangle$, $\bar k
\equiv \langle\bar k_\lambda \rangle$. Then 
Eq.~(\ref{mean_Q_explicit}) simplifies
\begin{eqnarray} 
  \langle Q_{e e}(E)\rangle &\simeq& \hbar {1 \over D} 
  {k_e \langle\gamma_{\lambda e}^2\rangle \over n \bar k 
   \langle{\rm var}(\gamma_\lambda)\rangle } \\
                            &\simeq& {\hbar \over \Gamma} 
                                     {k_e \over \bar k}
\label{overlapping_mean_delay}
\end{eqnarray} 
which tends to $0$ as $k_e \sim \sqrt\epsilon$.  The form of
Eq.~(\ref{overlapping_mean_delay}) and all the steps leading up to it
remain valid whether the Lorentzians are overlapping or not, as long
as the $\Delta E = \Delta\epsilon$ interval which we are averaging
over includes many of them.  

\subsection{On an isolated resonance}

If the target is cold enough that the
resonances are isolated, then as the incident particle's energy
$\epsilon \rightarrow 0$, adhering to the condition $\Delta\epsilon <
\epsilon$ will eventually result in $\Delta\epsilon$ becoming narrower
than the resonance widths.  It becomes possible then for
$\Delta\epsilon$ to be centered right around a single isolated
resonance at $E_\lambda^{(r)}$. In this case $\langle Q_{e
  e}(E)\rangle$ is found simply by putting $E = E_\lambda^{(r)}$,
because the spectrum $|g(E)|^2$ is well approximated by $\delta(E -
E_\lambda^{(r)})$.  So
\begin{equation}
  \langle Q_{e e}(E) \rangle = {\hbar \Gamma_{\lambda e} \over
  \Gamma_\lambda^2 } = {\hbar \over \Gamma_\lambda} 
                       {\Gamma_{\lambda e} \over \Gamma_\lambda}
 = {\hbar \over \Gamma_\lambda} {k_e \over n \bar k}\ .
\label{isolated_mean_delay}
\end{equation}
Even in this case there is the $\sqrt\epsilon$ behavior as $\epsilon
\rightarrow 0$ and there is no sticking.  

In the extreme case that there are no other open channels at all $(n =
1)$, $\langle Q_{e e}(E) \rangle \simeq {\hbar \Gamma_{\lambda e}
  \over \Gamma_\lambda^2} = {\hbar \over \Gamma_{\lambda e}}$ because
$\Gamma_\lambda = \Gamma_{\lambda e}$.  In fact, $e = 1$, and $\langle
Q_{e e}(E) \rangle$ diverges,
implying in this case that it
is possible to have the particle stick.  This is an exception to all
the cases above but is experimentally not so relevant because we may
always expect to find some exothermic channels open for a target with
many degrees of freedom.

\section{Inelastic cross sections and sticking}

Another physically motivated measure of the sticking probability may
be obtained by studying the total inelastic cross-section of the
collision.  The idea is that any long lived ``sticking'' is
overwhelmingly likely to result in an inelastic colision process; i.e.
that the scattering particle will leave in a different channel than it
entered with.  Using the original Wigner approach it is possible to
show that for our case where we have only one scattering degree of
freedom, the inelastic probability for an exothermic and endothermic
collision vanishes like $k_e$. The only possible exception to this is
a measure zero chance of a resonance exactly at the threshold energy,
$E^{\rm target}_e$.  In the event that there is a resonance
$E_\lambda^{(r)}$
close to but
above this threshold energy, it is only necessary that $E$ is below
$E_\lambda^{(r)}$ (by an energy of at least $\Delta E$, the spread in
  energy)
in order to observe the usual Wigner threshold behavior:
\begin{equation}
    P_{\rm inelastic} \rightarrow 0 \ \ {\rm like}\ \  k_e \propto
    \sqrt{\epsilon}
\end{equation} 
for the inelastic probability.
However our problem is unusual in the sense that because of the large
number of degrees of freedom of the target, we will always find
resonances between $E^{\rm target}_e $ and $E$ no matter how small
$E-E^{\rm target}_e = \epsilon$ is. Thus the Wigner regime is not
accessible. Still the
surprise is that a simple computation reveals the same behavior  holds  for large
$n$:
\begin{eqnarray}
    P_{\rm inelastic}(E) &=& \sum\limits_{c \not= e} P_{c
      \leftarrow e}(E) \\
       &=& \sum\limits_{c \not= e}|S_{c e}(E)|^2  \\
       &=& \sum\limits_{c \not= e}
           \sum\limits_{\lambda}\sum\limits_{\lambda'}
           {{\Gamma_{\lambda c}}^{1/2}{ \Gamma_{\lambda e}}^{1/2}
            \over E - E_\lambda^{(r)} -i \Gamma_\lambda /2}
           {{\Gamma_{\lambda' c}}^{1/2}{ \Gamma_{\lambda' e}}^{1/2}
            \over E - E_\lambda^{(r)} +i \Gamma_\lambda /2}\\
    \Rightarrow P_{\rm inelastic}(E)
       &=& \sum \limits_\lambda {\Gamma_\lambda \over (E - E_\lambda^{(r)})^2 +
       (\Gamma_\lambda /2)^2} \ \Gamma_{\lambda e}
\label{P_inelastic_of_E}
\end{eqnarray}
where we used the random sign property of the $\Gamma_{\lambda
  c}^{1/2}$'s and the understanding that  $\sum\limits_{c \not= e}
\Gamma_{\lambda c} \simeq \sum\limits_{\rm all \ c} \Gamma_{\lambda c}
= \Gamma_\lambda$. Since  the sum $\sum\limits_{c \not= e}$ is over the $n
  \gg 1$ 
open channels,
omission of a single term can hardly matter.
Apart from the factor $\hbar / \Gamma_\lambda$, the rhs of the above
equation is identical to the expression for $Q_{e e}(E)$ in
Eq.~(\ref{Q_ee}).  Averaging $P_{\rm inelastic}(E)$ over many
resonances $E_\lambda^{(r)}$ (overlapping or not) we may use the same
algebraic simplifications as before to show
\begin{equation}
  \langle P_{\rm inelastic} \rangle = {k_e \over \bar{k}}
\end{equation}
As $k_e$ tends to $0$, this gives the $\sqrt{\epsilon}$ Wigner
behavior showing that there is no sticking.

The above argument fails when there is only one open channel.  
There are no inelastic channels to speak
of.  In this case, if the energy $E$ coincides with a resonant energy
$E_\lambda^{(r)}$ we will have the exceptional case of sticking, as discussed at the end of the previous
section.  But as
pointed out there, this is primarily of theoretical interest only.

\section{Channel Decoherence}

The only case for which we stick is seen to be the case of when we are
sitting right on top of a resonance with the incoming energy so well
resolved  that we are completely within the resonance width, AND
there are no exothermic channels open. Having no such channels open amounts
to an infintesimally low energy for a large target.  Otherwise, the sticking
probability tends to 0 as $\sqrt{\epsilon}$ in every case.

\subsection{Time dependent picture}

  From the time independent point of view, the physical reason for the absence
of low energy sticking
is contained in the factor ${\Gamma_{\lambda e}\over \Gamma_\lambda}$
of Eq.~(\ref{Q_ee}).  This is the formation probability for the compound
state.  We will explain physically why it is small for $n \gg 1$.  The
resonance state is a many-body entangled state.  If we imagine the
decay of this compound state (already prepared by some other means
say) each open channel carries away some fraction of the outgoing
flux, with no preference for any one particular channel.  Running this
whole process in reverse it becomes evident that the optimum way to
{\em form} the compound state is to have each channel carry an
incoming flux with exactly the right amplitude and phase.  This
corresponds to however an entangled initial state.  With all the
incoming flux instead constrained to be in only one channel it becomes
clear that we are not exciting the resonance in the optimal way and
the buildup of amplitude inside is not so large; i.e., the compound
state has a small probability of forming.

The time dependent view is even more revealing.  Imagine a wave packet
incident on the system. For a single open channel Feshbach resonance,
the build-up of amplitude in the interior region can be decomposed as
follows.  As the leading edge of the wavepacket approaches the region
of attraction, most is turned away due to the quantum reflection
phenomena.  (It is a useful model to think of the quantum reflection
as due to a barrier located some distance away from the interaction
region.) The wavefunction in the interaction region constructively
interferes with new amplitude entering the region.  At the same time,
the amplitude leaving the region is out of phase with the reflected
wave, cancelling it and assisting more amplitude to enter.

Now suppose many channels are open.  All the flux entering the
interior must of course return, but it does so fragmented into all the
other open channels.  Only the fraction that makes it back into the
entrance channel has the opportunity to interfere (constructively)
with the rest of the entering wavepacket.  The constructive
interference is no longer efficient and is in fact almost negligible
for $n \gg 1$, thereby ruining the delicate process that was
responsible for the buildup of the wave function inside.  The
orthogonality of the other channels prevents interference in the
scattering dimension.  If we trace over the target coordinates,
leaving only the scattering coordinate, most of the coherence and the
constructive interference is lost, and no resonant buildup occurs.
Therefore, one way to understand the non-sticking is to say that decoherence is
to blame.

\subsection{Fabry-Perot and Measurement Analogy}

Suppose we have a resonant quantum mechanical Fabry-Perot cavity,
where the particle has a high probability of being found in between
the two reflecting barriers.  Now, during the time it takes for the
probability to build up in the interior, suppose we continually measure
the position of the particle inside.  In doing so we decohere the wave
function and in fact never find it there at all.  Alternatively,
imagine simply tilting one barrier (mirror) to make it non-parallel to
the first and redirecting the flux into an orthogonal direction, again
spoiling the resonance.  Measurement entangles other (orthogonal)
degrees of freedom with the one of interest, resulting in flux being
effectively re-directed into orthogonal states.  Thus the states of
the target (if potentially excitable) are in effect continually
monitoring (measuring) to see if the incoming particle has made it in
inside, ironically then preventing it from ever doing so.  The buildup
process of constructive interference in the interaction region,
described in the preceding paragraph, is slower than linear in $t$.
Therefore, the constant measurement of the particle's presence (and
resultant prevention of sticking) is an example of the Zeno
``paradox'' in measurement theory. 

\section{Conclusion}
We have presented a general approach to the low energy sticking 
problem, in the form of $R$-matrix theory.  This theory is well suited for the 
task, since it highlights the essential features of  multichannel scattering 
at low incident translational energy.  We did not need to make a harmonic 
or other approximate assumptions about the solid target, which is characterized by 
its long range interaction with the incoming particle and its density of states.
``Warm" surfaces are included in the formalism, and do not change the 
non-sticking conclusion.

Several supporting arguments for the non-sticking conclusion were given.
Perhaps most valuable is the physical decoherence picture associated with the 
conclusion that there is no sticking in the zero translational energy limit.  

Reviewing the observations leading up to the 
non-sticking conclusion, we start with the  near 100\% sticking 
in the zero translational energy limit classically (sticking
probability 1).  We then invoke 
the phenomenon of quantum reflection (Fig.\ref{fig:QR}), which  keeps the incident particle far from the
surface (sticking probability 0).  Third, we note that quantum reflection can be
overcome by resonances (Fig. \ref{fig:Feshbach}), and since resonances are ubiquitous in a many body target,
being the Feshbach states by which a partice could stick to the surface, perhaps
sticking approaches 1 after all.  Fourth, we suggest that decoherence 
(from the perspective of the incoming channel, with elestic scattering definded as
coherent) ruins the resonance effect, reinstating the quantum reflection as 
the determining effect.  Finally, then, there is no sticking, and the 
short answer as to why is: quantum reflection and many channel decoherence.
The ultrashort explanation is simply quantum reflection, but this is dangerous
and non-rigorous, as we have tried to show.

All this does not tell us much about how sticking turns on as incident 
translational energy is raised.  This is the subject of the following paper,
where a WKB analysis proves very useful.  Quantum reflection is a 
physical phenomenon liked directly to the failure of the WKB approximation.

\acknowledgements
This work was supported by the National Science Foundation
through a grant for the Institute for Theoretical Atomic and
Molecular Physics at Harvard University and Smithsonian
Astrophysical Observatory:National Science Foundation Award Number
CHE-0073544.

\appendix
\section{$\Gamma \simeq n D$}
\label{Bohr_Wheeler}
With the large number of degrees of freedom involved and assuming
thorough phase space mixing associated with the resonance we may
reasonably describe the compound state wavefunction by a classical
ensemble of points $(x, p_x, u, p_u)$ in the combined phase space of
the joint system given by the normalized distribution
\begin{equation}
  {1 \over \rho_C (E)} \delta(E - H(x, p_x, u, p_u)) .
\label{}
\end{equation}
It is understood in the above that the system is restricted to be in
the region $x<a$.  This makes all accessible states of energy $E$ with $x < a$
equally likely.  Then the rate of escape $\Gamma / \hbar$ through the
hypersurface $x = a$ of the members of this ensemble is
\begin{equation}
  {\Gamma \over \hbar}={1 \over \rho_C (E)} \int\limits_{x=a} du dp_u 
   \int\limits_{p_x \in [0,\infty]} dp_x {p_x \over m} 
   \delta(E - H(x, p_x, u, p_u)) .
\label{phase_space_int}
\end{equation}
$p_x / m$ is just the velocity in phase space of a point at $x = a$ in
the $\hat x$ direction.  At $x = a$ we have supposed no interaction.
Hence the Hamiltonian separates in Eq.~(\ref{phase_space_int}).
Therefore
\begin{eqnarray} 
  {\Gamma \over \hbar}&=&{1 \over \rho_C (E)} \int du dp_u 
   \int\limits_0^\infty d\left({p_x^2 \over 2m}\right) 
   \delta\left(E - \left({p_x^2 \over 2m} + H^{\rm target}(u, p_u)\right)\right) \\
       &=& {1 \over \rho_C} \int\limits_{H^{\rm target}(u, p_u) < E} du dp_u \\
       &=& {1 \over \rho_C} \Omega_C 
   \simeq {1 \over 2\pi\hbar\rho_Q} \Omega_Q = {1 \over 2 \pi \hbar} n D .
\label{}
\end{eqnarray} 
Therefore ${\Gamma \over D} \simeq n$.  $\rho_Q$ ($\rho_C$) is the
quantum (classical) density of states (phase space volume) of the
joint system at energy E.  $\Omega_Q$ ($\Omega_C$) is the quantum
(classical) total number of states (total phase space volume) of only the
target below energy E.  We have used the correspondence between the
Classical and Quantum density of states.  ${1 / \rho_Q}$ is
identified with $D$, and the number of states of the target having
energy less that $E$ is just $n$, the number of open channels.  

\section{Inelastic probability with background}
We show here that the inelastic probabilities remain essentially
unaffected in magnitude with the presence of a background term in
the S-matrix. In the isolated case the addition of $b_{c c'}$ to an
inelastic element $S_{c c'}$ simply changes the Lorentzian profile of
$|S_{c c'}|^2$. In the more important overlapping case, the energy
variation of $S_{c c'}$ is smooth in any case without background and
\begin{eqnarray}
|\bS_{c c'}|^2 &=&
\left|    B_{c c'} -i\sum\limits_{\lambda}{ \Gamma_{\lambda c}^{1/2}
          \Gamma_{\lambda c'}^{1/2}
          \over E_\lambda^{(r)} - E - i \Gamma_\lambda / 2}
\right|^2.\\
              &=& |B_{c c'}|^2+\sum\limits_{\lambda}{ \Gamma_{\lambda c}
          \Gamma_{\lambda c'}
          \over {(E_\lambda^{(r)} - E)}^2 + {\Gamma_\lambda}^2/4}
\end{eqnarray}
where we have used the random sign property of the products 
${ \Gamma_{\lambda c}^{1/2} {\Gamma_{\lambda c}}}^{1/2}$ to neglect the
2nd cross-term in comparison to the last one where again the same
property is used to simplify the double sum to a single one.  Summing
over all the inelastic channels then leads to the same result of Eq.~(
\ref{P_inelastic_of_E}) with an added term of 
$\sum\limits_{c\not=e} |B_{c c'}|^2$ which itself is proportional to 
$k_e$ as discussed at the end of Section \ref{overlapping_resonances}.


\begin{references}
\bibitem{lj} J. E.  Lennard-Jones {\it et. al.}, {\it Proc. R. Soc. }
London, Ser. A  {\bf 156} 6, (1936); Ser. A {\bf 156} 36, (1936).
\bibitem{wal} T.W. Hijmans, J.T.M. Walraven, and G.V.
Shlyapnikov, {\it Phys. Rev. } B {\bf 45} , 2561 (1992).
\bibitem{Brenig} W. Brenig, {\it Z. Phys.} B {\bf 36}, 227 (1980).
\bibitem{kohn} D.P.  Clougherty and W. Kohn, {\it Phys. Rev. } B, {\bf 46}
4921 (1992).
\bibitem{Bitt} E. R. Bittner, {\it J.  Chem. Phys.} {\bf 100}, 5314 (1993).
\bibitem{jul} P. S. Julienne and F. H. Mies, {\it J. Opt. Soc. Am.} {\bf B 6},
2257 (1989).
\bibitem{jul2} P. S. Julienne, A.M. Smith, and K. Burnett, Adv. At. Mol. Op. Phys.
{\bf 30}, 141 (1992).
\bibitem{landau:lifshitz} L.D. Landau and E.M. Lifshitz, {\it Quantum
   Mechanics (Non-relativistic Theory)} (Pergamon Press, Oxford (UK) 1981).
\bibitem{joachain:qct} C.J. Joachain, {\it Quantum Collision Theory}
(North-Holland, Amsterdam 1975).
\bibitem{flam} G.F. Gribakin and V. V. Flambaum, {\it Phys. Rev.} A
{\bf 48} 546 (1993).
\bibitem{cote}  R. C\^ote\', E. J. Heller, and A. Dalgarno,  ``Quantum suppression of cold atom collisions'' 
\pra  {\bf 53}, 234-41 (1996).
\bibitem{doyle}I. A. Yu, J.  Doyle, J. C. Sandberg,  C. L. Cesar, D.
Kleppner, and T. J. Greytak, {\it Phys. Rev. Lett} {\bf 71} 1589 (1993).
\bibitem{hesuper} J.  Doyle, J. C. Sandberg, I. A. Yu, C. L. Cesar, D.
Kleppner, and T. J. Greytak, {\it Phys. Rev. Lett} {\bf 67} 603 (1991);
C. Carraro and M.W. Cole,  {\it Phys. Rev. } B {\bf 45} , 12931 (1992);
 T.W. Hijmans, J.T.M. Walraven, and G.V.
Shlyapnikov, {\it Phys. Rev. } B {\bf 45} , 2561 (1992).
\bibitem{smith} F. T. Smith, {\it Phys. Rev. } {\bf 118}, 349 (1960).
\bibitem{bohr_wheeler} N. Bohr and J. Wheeler,  {\it Phys. Rev.} 
{\bf56} (5),416-450 (1939). see p. 426 Sec. III.
\end{references}
\end{document}